# Origin of negative differential resistance in molecular junctions of Rose Bengal

Anirban Bandyopadhyay[1]* and Y. Wakayama[2]

*[1]International Center for Young Scientists, National Institute of Material Science, 1-1 Namiki, Tsukuba, Ibaraki, 305-0044 Japan.

[2]Nanoassembly Group, National Institute for Materials Science, 1-1 Namiki, Tsukuba, Ibaraki, 305-0044 Japan.

Abstract: Negative differential resistance (NDR) is tuned at junctions of electronically different dimer and trimer of Rose Bengal on an atomic flat gold (111) surface. Isolated molecule did not show any NDR. But it was induced to show double NDR with large peak to valley ratio (1.8~3.1) in room temperature via charging its neighbor reproducibly by an electrical pulse. In some sections of junction by applying pulse one could destroy the phenomenon or regenerate it by STM manipulation of molecules. NDR was also independent of polaronic nature. It was possible to write bits 1 and 0 for cationic NDR (in dimer) and 00, 01, 10, 11 for di-anionic NDR (trimer) which generated 2/4 bit memory in a atomic scale junction showing importance of junction electronics in future of moletronics.

Corresponding author e-mail: anirban.bandyo@rediffmail.com

In the field of molecular electronics the conductance switches are one of the most important achievements that have been realized in the last decade though the advancement began long ago in the bulk inorganic and organic devices. The discovery of negative differential resistance by Esaki (1957) in the tunnel diode and McGinnes et al (1972)[1] in the organic molecule melanin was the first breakthrough in the respective field. Since the realization of a single molecule NDR by Tour (1999),[2] it is realized in several systems of the nano-materials world, beginning from nano dots,[3] bucky ball,[4] plastic nano-composites,[5] nanowires,[6] single wall nano tube (SWNT)[7] to even in carbon fiber interfaces.[8] In a single molecule characterization an orbital overlap with the external atomic electrodes is hardly reproducible, even the neighbors might move into the atom probe contact and modify the properties. A single unit is self-assembled in a monolayer or in a nanogap for the investigation.[9-10] In both the cases the electronic property of the substrate-molecule-STM tip system depends on the surface area of each unit exposed to the atomic tip. To be specific the vertically projected area on the substrate beneath the tip facet has the highest field effect only around $2nm^2$ for a tip of $30^0$ solid angle,[11] beyond which the field decays very sharply. Now if the exposed molecular area is less than the highest field region (Fig. 1a) then an inevitable question arises about the role of the junction in generating the electronic properties, which has hardly been addressed till now.

In a nanoscale measurement the NDR could originate from several reasons other than the characterizing system itself. For an example it is well known that if an atomic flat substrate is impure by a few oxide atoms then their redox activity might generate an NDR; or if the local energy levels of an STM tip can trap the charge then the trapping and detrapping of the charge could generate a reproducible NDR, which appears similar to the molecular NDR.[12] Recently, a debate is mounting on the true mechanism of NDR. In case of Tour switch (Reed *et al*)[2] Taylor Brandbyge and Stokbro (TBS)[13] has revealed the role of the side group in the intermolecular interaction and in NDR, challenging the concept of charging followed by the localization and delocalization of orbitals proposed by Seminario *et al.* Fan F. R. *et al*[14] has suggested an inelastic or two step tunneling through the induced states, Cornil *et al* has proposed[15] a bias induced alignment of the molecular orbitals, Di-ventra *et al* has proposed[16] that the rotation of the nitro group is responsible instead of a ring rotation, Dutta *et al* has suggested[17] the role of the local density of states of an STM tip and the electrostatic potential profile of the system, Nitzam and Ratner has suggested[18] for the polaron coupling with a necessity of the redox center while Donhauser et al has proved[19] that there is no necessity of the redox group and explored the role of conformational changes. As the exposed area of a Tour switch is around $0.5nm^2$, in the high field region of $2nm^2$ between the tip and the substrate there could be several molecular junctions, and all of them would contribute to the NDR (Fig. 1a), thus it would be impossible to decide which proposal is the correct one. To resolve this mystery one needs (i) a system with a large molecular surface (when they are self-assembled on an atomic flat substrate) to minimize the degree of a neighboring impact in a particular part of the molecular structure (Fig.1a, radius 'd') and finally (ii) that molecular system should not show NDR in an isolated measurement.

In this letter, it is shown statistically how a molecular junction could be induced to evolute an NDR in a controlled way; the phenomenon is typical of a molecular NDR but strongly conformation dependent. A Xanthene dye namely Rose Bengal (RB) that did not show any NDR in an isolated measurement when a current voltage (*IV*) measurement was carried out by fixing the STM tip at the OH terminal on the molecules body (COOH terminated plane lying on the gold (111) substrate at an angle $40^0$), was forced to generate an NDR bringing another molecule from its environment. Applying a typical electrical pulse for the molecular manipulation, various multistable junctions were formed to check the statistical aspects of Nitzam et al proposed role of the polarons[18] and TBS proposed[13] role of conformation modulation in an NDR. The phenomenon is tailored by bringing one to four molecules to its neighbor and forming a range of junctions. There is reported example where a single C60 molecule could not show NDR unless it formed a dimer.[20]

In the present case the area accessible to the STM tip is much larger than the exposed area in a Tour switch (in an RB it is ~$2nm^2$ with a radius 'd' of Fig 1a, while in a Tour switch it is $0.5nm^2$; radius 'a' of Fig 1a). There is a region on the molecular structure (we define as **jone**) where a single molecular property is observed, in a typical region near the junction, NDR phenomenon is observed (*J* jone) and in a region very near to the junction any measurement initiated a destruction of the conformation (*D* jone). Investigating several junctions a clear boundary was set between the molecular-NDR (might be observed by fixing the STM tip on a specific functional groups of the molecular structure ~2nm$^2$) and junction-NDR (~0.3nm$^2$). Finally, a new concept of junction switch is established in which an intrinsic molecular property is not an essential.

An atomic flat gold (111) substrate is used to grow the self-assembled films of Rose Bengal. To get a large area atomic flat substrate, 150nm thick gold film deposited by an e-beam evaporation on a freshly cleaved Mica was Hydrogen annealed (100% H$_2$) in a 600$^o$C for 30min. and etched in DMF (di-methyl formamide). The two processes were repeated several times in a cycle. The substrate after a thorough investigation in the UHV-STM was taken out in the air and immediately dipped into the 1μM solutions of RB and kept there for 10-12 hours to grow 0.8-1.2 monolayer on the surface (typical for the DMF solvent). However the growth rate was varied with the solutions of different concentration. Then the substrates were kept in the same solution for a few hours for the removal of an overdose of molecules and finally the films were kept under an UHV condition (~1.3×10$^{-8}$ Torr) for a few days to get a stable molecular film. The detail of the experimental method is published elsewhere (Bandyopadhyay et al *J. Phys. Chem.* B, **2006**)

A homogeneous self-assembled surface was grown (in some sections on the gold surface) with a molecular density varying in between 15-4 molecules/25nm$^2$, and on a lowest density film we have manipulated molecules by playing with the feedback loop and the applied pulse. A SAW-tooth pulse was applied to bring two molecules together or to separate them. The magnitude of the SAW tooth pulses was always kept at less than 0.6V ($dV/dt$ is large as the change ΔV~0.6V was made in Δt~1μs) and the pulses were applied at the brightest point of the Chlorine plane to bring two molecules nearer and on the Iodine plane to separate them (In a Rose Bengal one plane is perpendicular to the other, see the supplementary material). The neighbors were brought at atomic distances apart (~0.3nm) for several times (~200 events) from different directions to create different conformations and finally we have generated a statistical database of occurrence. The *IV* spectrum were taken (at 300K) at both sides of the junction (both sides of *D* jone) and on the molecular surface; and image of conformation was taken before and after the measurement. The spectra with significant conformation changes between the initial and final image were immediately rejected and the deviation in the *IV* spectra between both sides was significantly low and therefore ignored.

In an *IV* measurement it was observed that in case of a dimer junction, the NDR occurs only at the positive bias direction (0.88V) and NDR was not observed far from the junction on any other part of the molecule (inset, Figure 1 (a), for an exact position please see the SI online). The normalized density of states $NDOS = (dI/dV)(I/V)$ did not match with the theoretically computed energy density distribution of a single molecule or a dimer of any combination of the neutral and polaronic species (possibilities were $RB_1RB_2$, $RB_1RB_2^-$, $RB_1RB_2^{2-}$, $RB_1RB_2^+$, $RB_1^+RB_2^-$, $RB_1^+RB_2^+$, $RB_1^{2-}RB_2^+$, $RB_1^-RB_2^-$, $RB_1^{2-}RB_2^-$ etc, for dimerization the release energy varied between 20kCal to 60kCal). The junction dependence of the NDR effect (peak to valley ratio, PVR varied between 4 to 8 for different conformations, Fig. 2) could not be explained with the change in the discrete energy levels, given that the conformation induced energy level change is significantly less in the present case. Theoretically 1eV change in the energy level increases the PVR by several orders in magnitude. Among the various possible orientations of two RB molecules (energy varied between +98 kCal to -110 kCal) the only conformation of Fig 2 (a) showed the NDR effect reproducibly. This conformation occurred ~36%, while the other conformation of fig 2 (b) was generated ~20% of the total attempts, and we observed that the pure cationic dimer was unstable. None of the cation or anion active sites (detected by comparing the theoretical and experimental STM images) showed any sign of NDR when measured at the central region of the molecules body.

As NDR evolves only at junctions hence junction could be a quantum well (Q-well) wherein pseudo energy levels are generated or becomes active only in presence of another molecule. Role of another molecule is deceptive, might be equivalent of a functional charge reservoir. So number of reservoir was increased bringing another molecule to estimate reflected change in Q well properties.

Another molecule is brought to its neighbor forming a trimer; which resulted in a much higher conductivity (5 times higher at 1.5V) and two peaks (Fig. 2) observed might be due to two electrons involved in generating consecutive NDR peaks at 1.98V, 2.41V and -1.94V, -2.32V with PVR 2.4 and 2.2. One significant difference from cationic single NDR is that in the present case one observes peaks at both sides of bias direction and similarity remains in the survival of molecular property on some part of molecules body.

NDR is observed at all junctions between the three molecules and changing conformation does not change the position of peaks in *IV* spectrum but height (PVR changes significantly from 1.8 to 3.1). Third molecule taken away forcefully and the effect of double NDR was diminished regenerating single NDR effect. Based on ionization Shanzer *et al* have shown distinctively[21] existence of cationic (partially stable in dimer with neutral species, well stable in dimer with anionic species) single and double NDR forms; these ionization species significantly different from quinoidal reductions reported elsewhere.[22] One reason why single or dimer RB cannot generate double NDR is that molecule works as an initiator generating required local energy levels at the junction. And conformation modifies trap energy levels, which determines effective condition for inducing and diminishing the NDR phenomenon. Linear height profile across molecules shows a nanogap of ~0.3-0.4 nm. A nanogap is generally a well-defined polaron-neutral conformations varying energy -57 kCal to -151 kCal in case of dimer. Released energy in trimer varied between 3kCal to 50kCal except cationic trimer which release more than 100kCal. Change in frontier orbitals due to the trimerisation varies between stabilization of HOMO and LUMO depending on polaronic nature. More than three reservoirs do not provide any new phenomenon (behaves same as trimer, pseudo level might fail to provide more gating channel) hence it seems challenge of Ricca et al[23] to TBS[13] for conformation independence might not always be true.

In area average of *IV* spectrum in 50×50 nm$^2$ area in case of ultradense films contains NDR peaks but in case of the molecules largely separated isolated molecular features survive. Orientation of molecules is determined by matching minimum energy models of *RB, RB$^-$, RB$_2^{2-}$, RB$^+$* with STM image in MAYA which also reveled preference of surface order. Thus computed DOS of dimer and trimer did not match with NDOS features measured experimentally which proved energy levels near junctions are not originated from pristine molecule or polarons only. Also oxidation and reduction energies (less than 1 eV) do not match with NDR peaks as externally applied energy is spent in transition to equilibrium conformations apart from generating polarons. *IV* measurement on *J* jone i.e. either side *D* jone appeared similar irrespective of cationic, anionic, di-anionic or neutral species taking part in junction formation which proved that near junction irrespective of electronic feature both interacting systems behaves as single unit.

All these experimental and theoretical findings lead to the conclusion that, a unique quantum well is formed in junction whose pseudo-stationary states are controlled by charging of the molecules.

Semi-empirical computation using MOPAC showed that three RB molecules $RB_1$, $RB_2$ and $RB_3$, forming a trimer (or dimer) generated similar symmetry for equal effective charge minimum conformation which is unique feature of these system. If $RB_2$ is reduced by some means and have formed slightly destabilized $RB_1RB_2^-RB_3$ configuration. Co-ordination spheres around molecules increases decreases or remains unchanged during intermolecular charge transport or reaching equilibrium conformation. Role of co-ordination charge modulation on localized tunneling phenomenon is important as it tunes pseudo energy levels to generate 'gating' like control. Via direct or in-direct electron transfer pseudo-stationary states are retained typically in the quantum well. Starting from $RB_1^-RB_2RB_3$, one could obtain $RB_1RB_2^-RB_3$ or $RB_1RB_2RB_3^-$ via direct or superexchange.[24-25] From Landau Zener equation, probability of such an event is $P = 1 - \exp\left(\frac{-8\pi H_{ij}^2}{v|S_1 - S_2|}\right)$ where $P$ is the probability of electron transfer; $2H_{ij}$ is the gap between the energy surfaces $v$ the velocity of the nuclei and $S_1$-$S_2$ the difference in the slope at the crossing point between the surfaces without interaction. Theoretical quasi-classical proposal of tunneling by Pschenichnov's approach[26] in short periodic array suggests tunneling could be turned on or off by (gating) changing one or more of the well heights or depths. As cationic anionic and di-anionic species of RB has distinct absorption band, hence for dimer and trimer there should be two and three pseudo-stationary states respectively (difference in frontier orbitals of different RB species is reflected in PVR as it is directly related to change in energy with electron exchange). Trimer formation occurs with a few unique geometries (several saddle points of 10 different classes, $RB_1RB_2RB_3$, $RB_1RB_2^-RB_3$, $RB_1RB_2^{2-}RB_3$, $RB_1RB_2^+RB_3$, $RB_1^+RB_2RB_3^-$, $RB_1^+RB_2RB_3^{2-}$, $RB_1^-RB_2RB_3^{2-}$, $RB_1^-RB_2RB_3^-$, $RB_1^-RB_2^-RB_3^-$, $RB_1^+RB_2^+RB_3^+$) but only few of them (Fig. 2) were selected matching with the lowest two minimum energy conformations obtained theoretically. Comparing statistically most probable conformations of all above possibilities with experiment it was observed that Fig 2.(g) event has the highest probability for occurrence under random molecular manipulations.

Hence double NDR could be fluctuation of co-ordination spheres modulating pseudo energy levels in a typical Q well which is generated by coupling of strong dimer or trimer via H bonding and strong dipole-dipole interaction (dipole moments of *RB*, *RB*$^-$, *RB*$^{2-}$, *RB*$^+$ are 5.14D, 11.41D, 9.83D, 2.10D respectively). In case of dimers pseudo energy levels were less in number which might not have allowed double NDR to be evoluted but it still remains mystery how exactly junction conductivity and tunneling co-existed in harmony in the way experiment revealed.

Finally "write-read-erase-read" sequence[22] is applied on the junction to test transition between binary bit 1 and 0 for single cationic NDR and transition between 00, 01, 10, 11 states were studied for double anionic NDR effect. Hence junction dependent RAM and ROM application were realized, which might not be intrinsically molecular, but several different molecular geometries might have been involved to realize this phenomenon.

In conclusion how NDR could be tuned in the junctions of Rose Bengal molecule is shown in this letter. By changing different conformations and studying several cases of large database one could understand different important aspects of NDR. It was observed that any kind of redox activity might be essential to show typical NDR effect for a system (as Nitzam and Ratner[18] calculated) but conformation of that molecule has to play a significant role in explicit evolution of the phenomenon. In fact the role of its neighbor is to generate unique pseudo energy levels, which acts as gating like channels in vertical tunneling measurement generating "on" and "off" states (comparing TBS[13], neighbors are important but for different reasons).

Supporting information available online.

Figure Captions:

1. (a) Schematic view of NDR cases with molecular diameter a, b, c, d where semitransparent cylinder of diameter c represents nearly homogeneous field generated by atomic probe electrodes which dies out very sharply beyond diameter c; hence several junction will be active during characterization of molecules diameter a, and in case of molecule d when we focus our atomic electrode at junction we get a vulnerable to electric field region we named *D* jone, and beyond this region sections on molecules body within field cylinder is the *J* jone, where junction property is revealed. (b) Pschenichnov's quantum well in short periodic array, which have its own stationary state, and in addition to this a few pseudostationary states depending on the number of wells in periodic array, in case well not periodic immediate change would be in height and width of the wells which is expected in molecular cases. (c) Three co-ordination spheres around three RB molecules, arrows directed towards overlapping regions of co-ordination sphere.

2. 2D and 3D (inset) STM image of RB molecules whose junctions are shown by arrow, most probable equivalent energy minimized structure (determined by comparing STM image, and theoretical surface charge density, di-hedral angle and MAYA software) is also shown in inset. The STM images of combinations (height profile see online) are (a) Neutral-cation dimer 1, (b) Neutral-cation dimer 2, (c) Neutral-anion dimer, (d) Neutral-di-anion dimer, (e) Neutral-di-anion neutral trimer, (f) Anion-cation dimer, (g) Anion-neutral-di-anion trimer, (h) Neutral-neutral-neutral trimer, (i) Rose Bengal molecular structure and its three possibilities of forming a polaron, cationic A, anionic B and di-anionic when B and C occurs together.

3. Current voltage measurement (300K) on cation-neutral dimer (a) and anion-neutral-dianion trimer (b) of Rose Bengal whose corresponding images are given in Fig. 2(a) and Fig. 2(g). Inset we can find differential spectrum for both cases indicating binary and tetranary bits and height profile of quantum well (for scaling see online). PVR values of single NDR is 6, double NDR is 2.48 and 2.26 while Tour switch[2] is 1.5.

Figure 1.

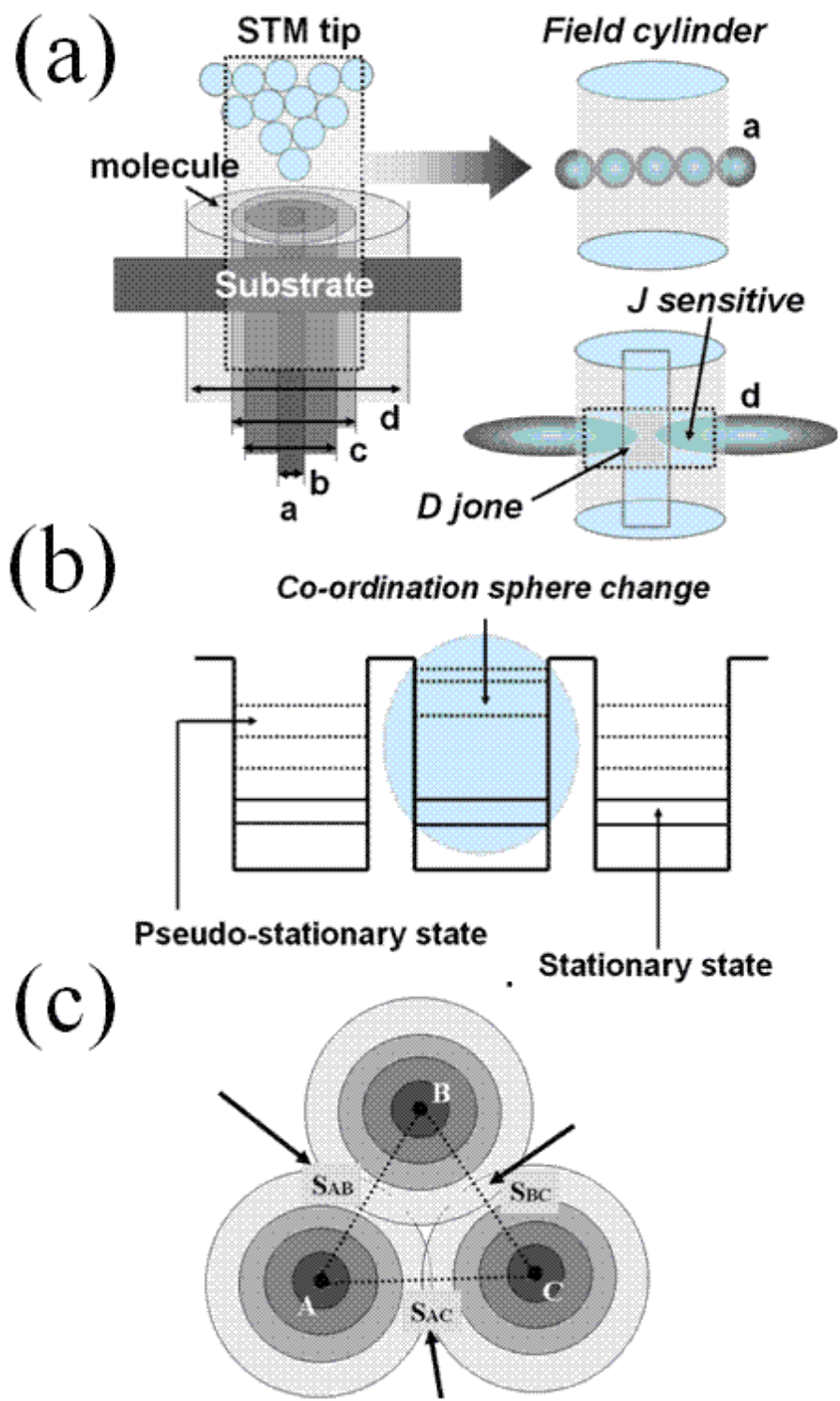

Figure 2.

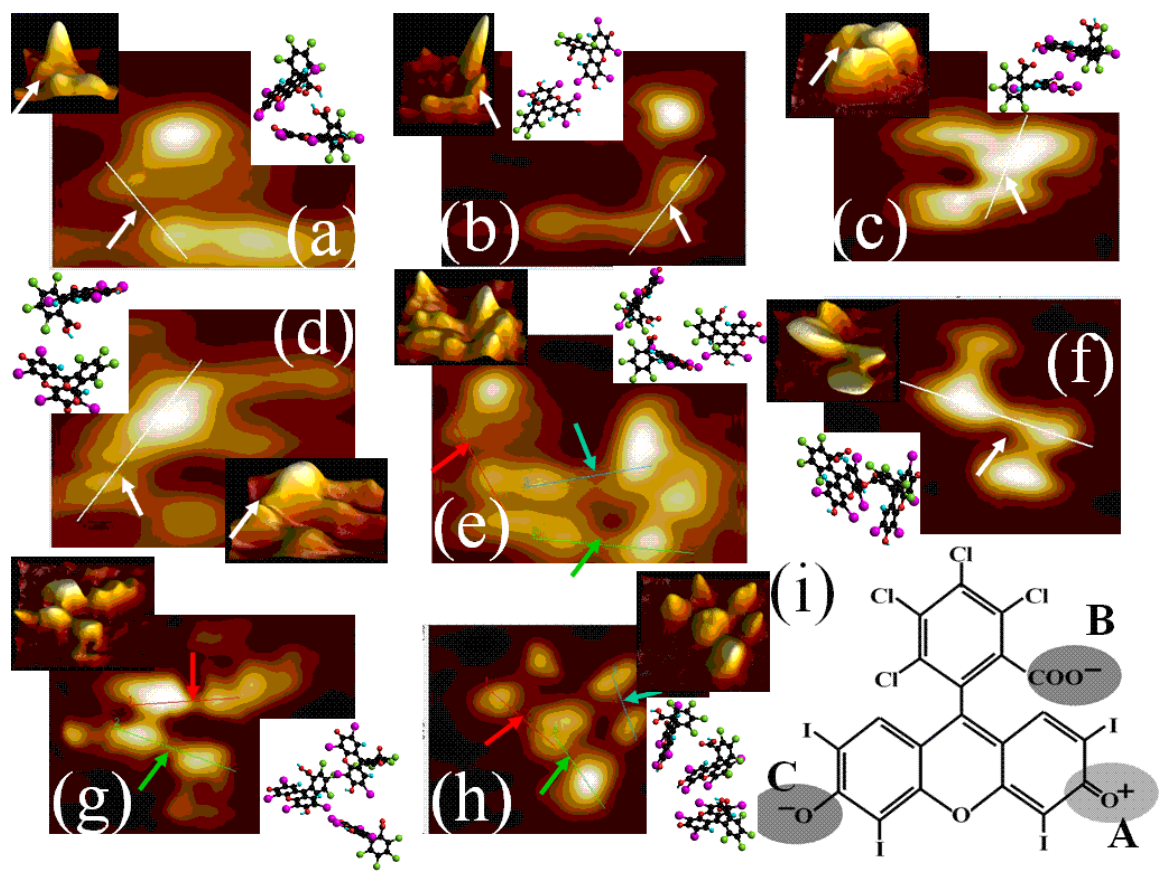

Figure 3.

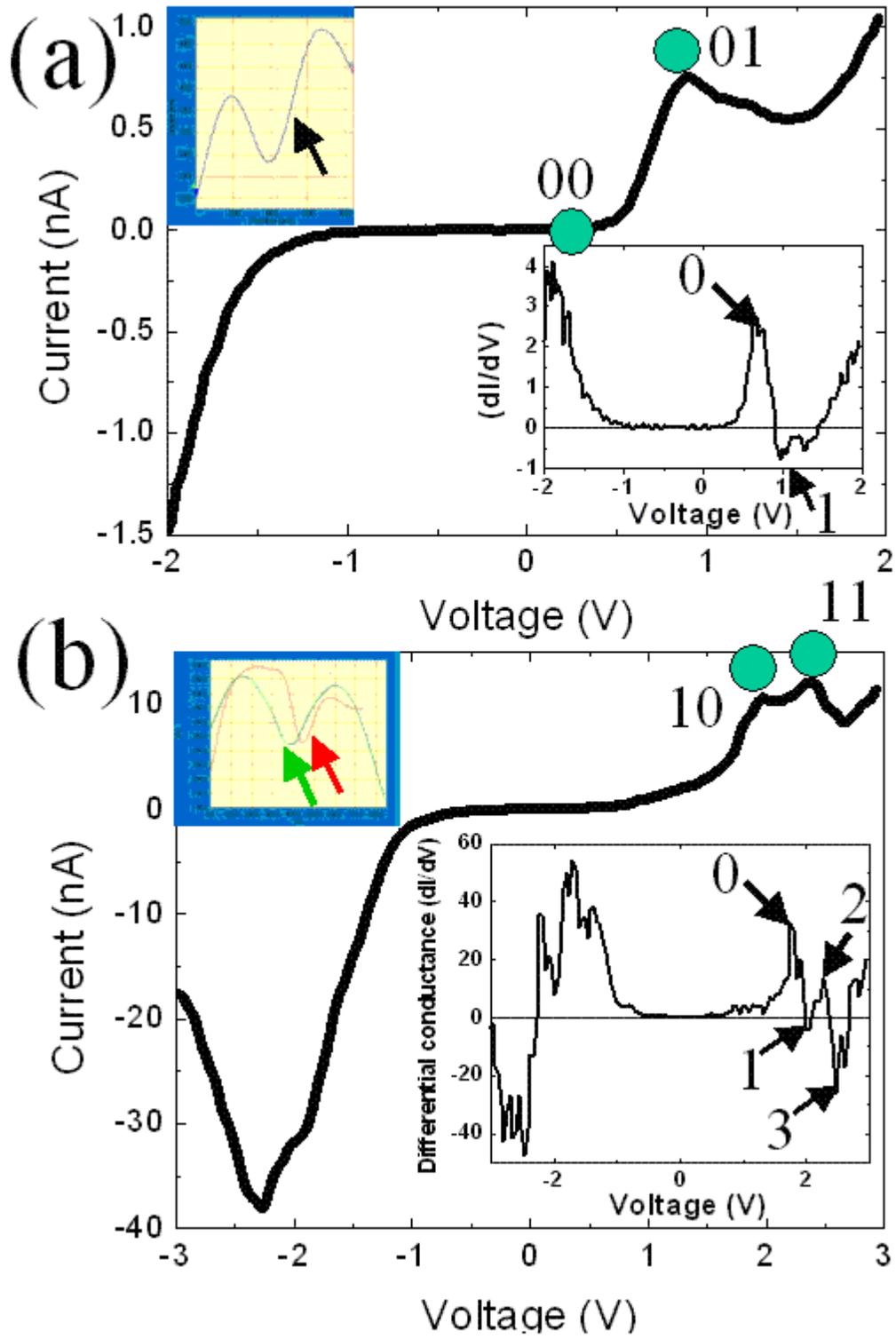